\documentclass[twocolumn,pra,psfig,showpacs,superscriptaddress]{revtex4}

\usepackage{graphicx}
\usepackage{times}

\newcommand{\be}{\begin{equation}}
\newcommand{\ee}{\end{equation}}
\newcommand{\ben}{\begin{eqnarray}}
\newcommand{\een}{\end{eqnarray}}

\begin{document}

\title{Maximum Entanglement in Squeezed Boson and Fermion States}

\author{F. C. Khanna}
\affiliation{Theoretical Physics Institute, University of Alberta,
Edmonton, Alberta T6G 2J1, Canada}
\affiliation{TRIUMF, Vancouver,
British Columbia V6T 2A3, Canada}
\author{J. M. C. Malbouisson}
\affiliation{Instituto de F\'{\i}sica, Universidade Federal da
Bahia, 40210-340, Salvador, BA, Brazil}
\affiliation{Theoretical
Physics Institute, University of Alberta, Edmonton, Alberta T6G 2J1,
Canada}
\author{A. E. Santana}
\affiliation{Instituto de F\'{\i}sica, Universidade de
Bras\'{\i}lia, 70910-900, Bras\'\i lia, DF, Brazil}
\affiliation{Theoretical Physics Institute, University of Alberta,
Edmonton, Alberta T6G 2J1, Canada}
\author{E. S. Santos}
\affiliation{Centro Federal de Educa\c c\~ao Tecnol\'ogica da Bahia,
40030-010, Salvador, BA, Brazil}

\begin{abstract}
A class  of squeezed boson and fermion states is studied with
particular emphasis on the nature of entanglement. We first
investigate the case of bosons, considering two-mode squeezed
states. Then we construct the fermion  version  to show that such
states are maximum entangled, for both  bosons and fermions. To
achieve these results, we demonstrate some  relations involving
squeezed boson states. The generalization to the case of fermions is
made by using Grassmann variables.
\end{abstract}

\pacs{42.50.Ar; 42.50.Dv; 03.67.Lx}

\maketitle

Boson coherent states have played an important role in the
production and detection of Bose-Einstein condensate. The
experimental tools of laser cooling, magnetic and magneto-optic
traps have advanced tremendously~\cite{fer1,fer2}. This has lead to
a consideration of producing a degenerate Fermi gas as well as
condensates of rare isotopes, a fact that has been achieved
experimentally~\cite{fer2,fer3,fer4,fer5,fer6}. These results
suggest that a careful study of the Fermi coherent state and Fermi
density operators  is needed. Some of these concepts have been
introduced for fermion systems long ago~\cite{fer7,fer8,fer9,fer10},
but only recently, Cahill and Glauber~\cite{fer11} have discussed
notions such as $P$-function, $Q$-function and Wigner function for
fermions;  all of them are described as a counterpart of the Boson
systems and are made possible through the use of Grassmann
variables.

In addition to the idea of achieving a Fermi degenerate gas, there
is a great deal of interest in studying entangled states of
multipartite systems~\cite{fer12,fer121}. The existence of entangled
states is directly related to the nature of the quantum mechanics
formalism, based on the structure of Hilbert space and the
superposition principle. The present interest in  entangled states
is strongly driven by the understanding that this is the focal point
of studies  leading to teleporting of quantum states, from one locus
to another, which is also the basic ingredient of quantum
computers~\cite{tel1,tel2}. The necessity to study such entangled
states is greatly increased with the suggestion that the conditions
for teleporting, however, require specific states characterized by
maximum entanglement. Measure of entanglement has been discussed in
the literature in different
ways~\cite{fer13,fer14,fer15,fer16,fer17,fer171}, and the
identification and construction of maximum entangled states have
also been addressed~\cite{pli1,gras1}.

Consider the state $| \psi \rangle_{ab} $ of a bipartite system
$(A,B)$. Define the reduced density operator $\rho _{a}$, by
\begin{equation}
\rho _{a}={\rm Tr}_b ( |\psi \rangle_{ab} \,_{ba}\langle \psi |),
\label{reden1}
\end{equation}
where ${\rm Tr}_b$ stands for the trace over the variables of the
subsystem $B$. A measure of the amount of entanglement of the state
$ |\psi \rangle_{ab}$ is given by the von Neumann entropy associated
with the reduced density operator~\cite{pli1}, ${\cal S}_{a}=-{\rm
Tr} (\rho _{a}\ln \rho _{a})$. The question addressed in this paper
is concerned with the maximum entangled state within the set of all
pure states having a given (fixed) reduced energy.

The entropy ${\cal S}_{a}$ is a homogeneous function of first degree
in its dependency on ${\cal E}_{a}$, the energy of the system $A$.
Then since we wish to maximize ${\cal S}_{a}[\rho _{a}]$ we require
$\delta {\cal S}_{a}[\rho_{a}]=0$ under the constraints
\begin{equation}
{\cal E}_{a}=\langle {H}_a\,\rangle ={\rm Tr}(\rho _{a}{H}_a) ,
\,\,\,\,
 {\rm Tr}\rho _{a}=1,
\label{cons1}
\end{equation}
where $H_a$ is the energy operator of the $A$-component. Following
standard procedures, we derive then a constraint equation for $\rho
_{a},$ that is $\lambda_0 - 1 + \lambda_1 H_a - \ln \rho_a=0, $
where $\lambda_0$ and $\lambda_1$ are Lagrange multipliers. Solving
this equation, we get a Gibbs-like density operator, that is
\begin{equation}
\rho _{a}=\frac 1Z\exp (\lambda_1{H}_a),  \label{cons3}
\end{equation}
where $Z=\exp (1-\lambda_0).$ Multiplying the constraint equation by
$\rho _{a}$, taking the trace and using Eqs.~(\ref{cons1}) and
(\ref{cons3}), we get $ \ln Z=\lambda_1 {\cal E}_{a}+{\cal S}_{a}$.

For the sake of convenience, let us write $\lambda_1=-\tau $, then
we have $ -\tau^{-1} \ln Z = {\cal E}_{a} - \tau^{-1} {\cal S}_{a}$.
The function $ F(\tau )=-\tau^{-1} \ln Z $  describes the Legendre
transform of ${\cal S}_{a}$ since we assume that $\tau =\partial
{\cal S}_{a}/\partial {\cal E}_{a}.$ Here $\tau $ is an intensive
parameter describing the fact that the energy average ${\cal
E}_{a}=\langle{H}_a\rangle $, given by Eq.~(\ref{cons1}), is
constant in the state described by $\rho _{a}$. Therefore, a state
$| \psi \rangle_{ab} $ with reduced energy ${\cal E}_{a}$ is a
maximum entangled state when the corresponding reduced density
matrix, defined in Eq.~(\ref{reden1}), is written as a canonical
Gibbs-ensemble distribution, explicitly given by Eq.~(\ref{cons3}),
such that $ Z=[1-\exp (-\tau )]^{-1}$. Using this approach, examples
of maximum entanglement states were explicitly constructed in
\cite{gras1}, considering as a guide the thermofield dynamics (TFD)
formalism.

In the present work we extend that analysis to show that two-mode
squeezed states (TMSS) are maximum entangled, within a class of
states with fixed energy, for both fermions and bosons. We first
investigate the case of bosons and, then, we construct the fermion
version, extending a preliminary study by Chaturvedi et
al.~\cite{fer10}. For the case of fermions, the situation is more
intricate, demanding the notion of coherent fermion state and
density operator, which is achieved by using Grassmann variables. In
any case we explore the similarities among these squeezed states
with those found in the TFD formalism~\cite{ume1,ume2,ume3}.

In order to analyze entanglement, we derive some general properties
of the usual Caves-Schumaker (CS) state for bosons~\cite{cv1}.
Consider a two-boson system specified by the operators $a$ and $b$
obeying the algebra $[a,a^{\dagger}]=[b,b^{\dagger}]=1,$ $[a,b]=0$,
with unitary displacement operators $D_a(\xi)=\exp[\xi a^{\dagger} -
a\xi ^{\ast}]$ and $ D_b(\eta)=\exp[\eta b^{\dagger} - b\eta
^{\ast}]$, and define the two-mode squeezing operator
\begin{equation}
S_{ab}(\gamma)=\exp[\gamma (a^{\dagger}b^{\dagger} -
ab)],\label{sq2}
\end{equation}
with $\gamma$ being a real non-negative number for simplicity. Using
standard TFD formulas~\cite{ume1,ume2}, we get the useful relations
\begin{eqnarray*}
a(\gamma ) &=&S_{ab}(\gamma ) a S_{ab}^{\dagger }(\gamma )= u(\gamma
)a-v(\gamma
)b^{\dagger },\\
b(\gamma ) &=&S_{ab}(\gamma ) b S_{ab}^{\dagger }(\gamma )=u(\gamma
)b-v(\gamma )a^{\dagger },
\end{eqnarray*}
and the corresponding relations for $a^{\dagger }(\gamma) $ and
$b^{\dagger }(\gamma)$, where $u(\gamma)=\cosh \gamma$ and
$v(\gamma)=\sinh \gamma. $

First, consider the TMSS defined by $|\gamma\rangle_{ab} =
S_{ab}(\gamma ) | 0 \rangle_{ab}$, where $| 0 \rangle_{ab} =
|0\rangle_a \otimes 0\rangle_b \equiv |0\rangle_a
 |0\rangle_b$ is the two-mode vacuum such that
$a|0\rangle_a=b|0\rangle_b=0$. For this squeezed state, we have
$a(\gamma )|\gamma \rangle_{ab} = b(\gamma )|\gamma \rangle_{ab} =0
$. Another important result is that $S_{ab}(\gamma) $ is a canonical
transformation, that is, $[a(\gamma ),a^{\dagger}(\gamma
)]=[b(\gamma ),b^{\dagger}(\gamma )]=1$ and $[a(\gamma ),b(\gamma
)]=0$. Now, using the operator identity $\exp[\gamma(A+B)] =
\exp[(\tanh \gamma) B]\, \exp[(\ln\cosh \gamma)C]\, \exp[(\tanh
\gamma) A]$, with $A=-ab$, $B=a^{\dagger}b^{\dagger}$ and
$C=[A,B]=-a^{\dagger}a-bb^{\dagger}$, the TMSS can be written as
\begin{equation} \nonumber
|\gamma\rangle_{ab} = \frac{1}{\cosh\gamma}\exp[(\tanh
\gamma)a^{\dagger}b^{\dagger}]\,|0\rangle_{ab}.
\end{equation}
Changing the parametrization by taking $
\cosh\gamma=[1-\exp(-\tau)]^{-1/2}$, so that $\tanh \gamma =
\exp(-\tau/2)$, and defining $Z(\tau)=[1-\exp(-\tau)]^{-1}= {\rm
Tr}\exp[-\tau a^{\dagger}a]$, we find
\begin{equation}
|\gamma\rangle_{ab} = \frac{1}{\sqrt{Z(\tau)}}\,\sum_{n=0}^{\infty}
e^{-\tau n/2}|n\rangle_a|n \rangle_b.
\end{equation}
Therefore, the TMSS $|\gamma\rangle_{ab}$ can be written as
\begin{equation}
|\gamma\rangle_{ab}=\sqrt{f_{a}(\tau)}\,\sum_{n=0}^{\infty}
|n\rangle_a|n \rangle_b, \label{esd1}
\end{equation}
where $f_{a}(\tau)=\exp(-\tau a^{\dagger}a)/Z(\tau).$

The CS states are introduced by the application of $S_{ab}(\gamma)$
to a two-mode coherent state, that is $|\xi,
\eta,\gamma\rangle\rangle=S_{ab}(\gamma
)D_a(\xi)D_b(\eta)|0\rangle_{ab}$. We can show that $|\xi,
\eta,\gamma\rangle\rangle=|\bar{\xi}, \bar{\eta},\gamma\rangle$,
with $|\bar{\xi},
\bar{\eta},\gamma\rangle=D_a(\bar{\xi})D_b(\bar{\eta})S_{ab}(\gamma
)|0\rangle_{ab}$, where
\begin{equation}
\left(
\begin{array}{c}
\bar{\xi} \\
\bar{\eta}^{\ast }
\end{array}
\right) ={\cal B}_B(\gamma )\left(
\begin{array}{c}
\xi \\
\eta^{\ast }
\end{array}
\right) ,  \label{c2bog4}
\end{equation}
${\cal B}_B(\gamma) $ being the matrix form associated to
$S_{ab}(\gamma)$ given by
\begin{equation}
{\cal B}_B(\gamma )=\left(
\begin{array}{cc}
u(\gamma ) & -v(\gamma ) \\
-v(\gamma ) & \;\; u(\gamma )
\end{array}
\right) .  \label{c2bog5}
\end{equation}
For $\xi=\eta=0$, the CS state reduces to the TMSS $|\gamma
\rangle_{ab}=S_{ab}(\gamma)|0\rangle$, which has the same structure
as the thermal vacuum state used to introduce TFD~\cite{ume1,ume2}.
We explore this result to show that $|\xi,
\eta,\gamma\rangle\rangle$ can be used to define a Gibbs-like
density.

Following the prescription discussed before, taking $
\rho_{ab}=|\xi, \eta, \gamma\rangle\langle \gamma, \eta,\xi |$, we
have to perform the calculation of the reduced density matrix, say
$\rho_{a}={\rm Tr}_b\rho_{ab}$. Using the notation, $|r\rangle_b=
{n!}^{-1/2}(b^{\dagger})^r|0\rangle_b$ (similarly for mode $a$), we
can write the matrix elements $\langle s|\rho_{a}|t\rangle = \sum_r
{}_a\langle s| {}_b\langle r|\rho_{ab}|r\rangle_b|t\rangle_a$ as
\begin{eqnarray*}
\langle s|\rho_{a}|t\rangle&=&\sum_{r,m,n} {}_a\langle  s|
D_a(\xi)\sqrt{f_{a}(\tau)}|n\rangle_a{}_b\langle
r|D_b(\eta)|n\rangle_b\, \\
&&\times\; {}_b\langle m|D_b(\eta)^{\dagger}|r\rangle_b{}_a\langle
m|\sqrt{f_{a}(\tau)}D_a(\xi)^{\dagger}|t\rangle_a.
\end{eqnarray*}
Changing the order of the matrix elements in the $b$ mode, and using
the completeness relation, we obtain $\langle s|\rho_{a}|t\rangle =
\langle s|D_a(\xi)f_{a}(\tau)D_a(\xi)^{\dagger} |t\rangle $. Thus,
we get
\begin{equation}
\rho_{a} = D_a(\xi)f_{a}(\tau)D^{\dagger}_{a}(\xi) =
Z^{-1}(\tau)\exp (-\tau a^{\dagger}(\xi)a(\xi)),
\end{equation}
where $a(\xi) = D_a(\xi) a D_{a}^{\dagger}(\xi)$ is the displaced
annihilation operator; $\rho_{a}$ is, therefore, a Gibbs-like
density. In particular, for $\xi=0$, we find $\rho_a = f_{a}(\tau)$
showing that the TMSS $|\gamma\rangle_{ab}$ also generates a
Gibbs-like density.

Using the displaced Fock's basis, $\{ D_a(\xi)|n\rangle_a \}$, we
show that the reduced von Neumann entropy for a CS state is
\begin{equation}
{\cal S}(\tau) = \frac{\tau}{e^{\tau}-1}- \ln \left( 1 - e^{-\tau}
\right) .
\end{equation}
Thus, all CS states, with the same (fixed) squeezing parameter, have
the same amount of entanglement independent of the displacement
parameters. Among them, the one having the smallest energy is the
TMSS ($\xi = 0$); its reduced energy (${\cal E}_a = {\rm Tr}(\rho_a
a^{\dagger} a)$) is given by ${\cal E}(\tau) = \left( e^{\tau} -1
\right)^{-1}$. Since both actions of displacing and squeezing the
vacuum lead to states with greater energy, the TMSS is the maximum
entangled state when the energy is fixed.

Let us now analyze comparatively the amount of entanglement of the
TMSS. Consider the state $\left| \Psi \right\rangle_{ab}^{(N)} =
N^{-1} \sum_{n=0}^{N-1} \left| n \right\rangle_a \left| n
\right\rangle_b$ which has reduced energy and entropy given by
${\cal E}^{\prime}(N) = (N-1)/2$ and ${\cal S}^{\prime}(N) = \ln N$.
This state has the greatest amount of entanglement among all pure
states belonging to the finite ($N^2$) dimensional subspace spanned
by $\{ \left| 0 \right\rangle_a \left| 0 \right\rangle_b , \left| 0
\right\rangle_a \left| 1 \right\rangle_b , \dots , \left| N-1
\right\rangle_a \left| N-1 \right\rangle_b \}$, corresponding to
equal occupation probability. Naturally, as $N\rightarrow\infty$,
both energy and amount of entanglement of $\left| \Psi
\right\rangle_{ab}^{(N)}$ goes to $\infty$.

Now, take another parametrization of the TMSS by writing $\tau =
\ln(\chi+1)-\ln(\chi-1)$; the limit situations of zero and infinite
squeezing correspond to $\chi=1$ ($\gamma=0, \;\tau=\infty$) and
$\chi=\infty$ ($\gamma=\infty,\; \tau=0$), respectively. The reduced
energy and the amount of entanglement are then written as ${\cal
E}(\chi) = (\chi-1)/2$ and ${\cal S}(\chi) = \left[
(\chi+1)\ln(\chi+1) - (\chi-1)\ln(\chi-1) - 2\ln 2 \right]/2$. We
find that both ${\cal E}(\chi)$ and ${\cal S}(\chi)$ go to $\infty$
as $\chi\rightarrow\infty$, with ${\cal S}(\chi)\sim \ln \chi$ in
this limit. In Fig.~1, we plot the difference between ${\cal
S}(\chi)$ and $\ln \chi$, showing explicitly that the TMSS has an
amount of entanglement greater than that of the state $\left| \Psi
\right\rangle_{ab}^{(N)}$ with the same energy,  for all $N \geq 2$.

%%%%%%%%%%%%%
\begin{figure}[ht]
\begin{center}
\scalebox{0.60}{{\includegraphics{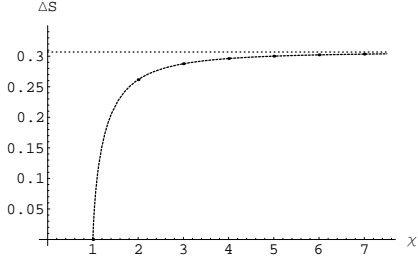}}}
\end{center}
\caption{Plot of $\Delta {\cal S} = {\cal S}(\chi) - \ln \chi$ as a
function of $\chi$.} \label{fig1}
\end{figure}
%%%%%%%%%%%%%

We now consider a similar analysis for fermions. A coherent fermion
state~\cite{fer7,fer8,fer11} can be defined by introducing a
displacement-like operator $D(\alpha)$, where $\alpha $ is a
Grassmann variable, trying to reproduce formally the basic results
of the boson case. This is accomplished  in the following way.
Consider $\alpha$  and $\beta$ two Grassmann numbers, then
$\{\alpha,\beta\}=\alpha\beta+\beta\alpha =0,$ and moreover,
$\{\alpha,a\}=\{\alpha,a^{\dagger}\}=0$, where the  fermion operator
$a$ and $a^{\dagger}$ satisfy the anticommutation relation
$\{a,a^{\dagger}\}=1.$ Notice that we maintain the same notation for
the creation and annihilation operators as that for bosons.  The
complex conjugation is an antilinear mapping $ *:\alpha \,
{\rightarrow} \, \alpha^*$ such that, for a general expression
involving Grassmann numbers and the operators $a$ and $a^{\dagger}$,
we have $(\alpha_i\alpha_j^*+c \beta_i\beta_j )^*
=\alpha_j\alpha_i^*+c^*\beta_j^*\beta_i^*; \,\, c\in {\bf C}$ and
$(a_i\alpha_ja_k^{\dagger }\beta_l^*)^{\dagger}= (\beta_l^*\alpha_j
a_ia_k^{\dagger })^{\dagger} = a_k a_i^{\dagger }\alpha_j^*
\beta_l$. The Grassmann variable $\alpha$ is considered independent
of $\alpha^*$.

The fermion displacement operator is defined by
\begin{equation}\label{disp1}
D_a(\alpha)=\exp( a^{\dagger}\alpha - \alpha^* a ),
\end{equation}
such that $D_a(\alpha)aD^{\dagger}_a(\alpha)=a-\alpha,$ with the coherent state
given by $|\alpha \rangle=D_a(\alpha)|0\rangle_a$ and $a|\alpha
\rangle=\alpha|\alpha
\rangle.$ We can then prove that the dual coherent state is given by
$\langle \alpha|=\langle 0|D^{\dagger}_a(\alpha),$ with
$D^{\dagger}_a(\alpha)=D^{-1}_a(\alpha)$. As a consequence
\[
\langle \alpha|\beta\rangle=\exp(\alpha^* \beta-\frac{1}{2}\alpha^* \alpha
-\frac{1}{2}\beta^* \beta)
\]
and $\langle \alpha|D^{\dagger}_a(\alpha)= \langle \alpha|\alpha^*$.
In terms of the number basis, the state $|\alpha \rangle$ is written
as
\begin{equation}
|\alpha \rangle  =  e^{- \alpha^* \alpha/2}
\sum_n(-\alpha)^n|n\rangle
\end{equation}
and, then, we have $ \langle n|\alpha\rangle =
\exp(-\alpha^*\alpha/2)(\alpha )^n .$

The integration is defined, as usual, by $\int d\alpha=0$ and $\int
d\alpha \alpha=1$. Note that, in particular, we have $\int d\alpha^*
\alpha^*=1$ (resulting in $(d\alpha)^*=-d\alpha^*$), $\int d\alpha^*
d\alpha \,\alpha \alpha^*  = 1$ and $\int d\alpha^* d\alpha \,
|\alpha\rangle\langle \alpha| = 1$.

Cahill and Glauber~\cite{fer11} introduced the following coherent
fermion state representation for a density operator, i.e. $\rho=\int
d^2 \alpha \, P(\alpha)\, |-\alpha\rangle\langle \alpha |$, where
$P(\alpha)$ is the corresponding $P$-function. Notice that the
density operator
\begin{equation}\label{cahill12}
\rho_{\alpha}=|-\alpha\rangle\langle \alpha |.
\end{equation}
possesses the expected properties to be taken as representing the
coherent state $|\alpha\rangle$. First, it is normalized, ${\rm
Tr}\rho_{\alpha}=1$. This property can be proved from the matrix
representation of $\rho_{\alpha}$, by calculating $\langle m
|\rho_{\alpha}|n\rangle=\exp(\alpha\alpha^*) (-\alpha)^m(\alpha^*)^n
$, giving rise to
\[
\rho_{\alpha}=\left(
\begin{array}
[c]{cc}
1-\alpha^* \alpha & \alpha^*  \\
-\alpha & \alpha^* \alpha
\end{array}
\right).
\]
Secondly, ${\rm Tr}(\rho_{\alpha}a^{\dagger}a)=\alpha^* \alpha$,
which is similar to the boson case. Observe that $\rho_{\alpha}, $
although not being hermitian, is introduced in such way that
$\rho^{\dagger}=\rho$~\cite{fer11}.

Using the properties  described above, we can prove that the
displaced fermion number state is given by
\[
D_a(\alpha)|n \rangle = (a^{\dagger} -\alpha^*)^n|\alpha\rangle ,
\]
with $n=0,1.$ Another property useful for calculations but
reflecting also the nature of $\rho_{\alpha} $ is given by
\begin{equation}
\langle m
|\rho_{\alpha}|n\rangle=(-1)^{m(n+1)} \langle \alpha| n \rangle\langle m|\alpha
\rangle ,
\end{equation}
where $\langle m |\rho_{\alpha}|n\rangle = \langle m |-\alpha\rangle
\langle \alpha |n\rangle $. Observe that for $m=n$ we have $\langle
n |-\alpha\rangle \langle \alpha |n\rangle = \langle \alpha| n
\rangle \langle n|\alpha \rangle $. The usefulness of this result
can be seen in the proof that ${\rm Tr}\rho_{\alpha}=1$.

Let us now consider a two-fermion system, specified by the operators
$a$ and $b$ satisfying the algebra
$\{a,a^{\dagger}\}=\{b,b^{\dagger}\}=1,$ with all the other
anticommutation relations being zero. A fermionic two-mode squeezed
vacuum state is defined by $|\gamma \rangle_{ab}=S_{ab}(\gamma
)|0\rangle_{ab}$, where $\gamma$ is still a real number and $
S_{ab}(\gamma)=\exp[\gamma (a^{\dagger}b^{\dagger} - ab)]. $ Some
useful formulas can be derived using $S_{ab}(\gamma )$, that is,
\begin{eqnarray*}
a(\gamma ) &=&S_{ab}(\gamma ) a S^{\dagger }_{ab}(\gamma )= u(\gamma
)a-v(\gamma
)b^{\dagger },\\
b(\gamma ) &=&S_{ab}(\gamma ) b S^{\dagger }_{ab}(\gamma )=u(\gamma
)b+v(\gamma )a^{\dagger },
\end{eqnarray*}
where now $u(\gamma)=\cos(\gamma) $ and $v(\gamma)=\sin(\gamma)$.
Thus, for the two-mode squeezed vacuum state $|\gamma \rangle_{ab}$
we have $a(\gamma )|\gamma \rangle_{ab} = b(\gamma )|\gamma
\rangle_{ab} =0$, since $a|0\rangle_a=b|0\rangle_b=0$.

The squeezing operator $S_{ab}(\gamma)$ is a canonical
transformation, in the sense that, as in the case of bosons,
$\{a(\gamma ),a(\gamma )^{\dagger}\}=\{b(\gamma ),b(\gamma
)^{\dagger}\}=1$ and $\{a(\gamma ),b(\gamma )\}=0$. The matrix form
${\cal B}_F(\gamma )$ associated to $S_{ab}(\gamma)$  is
\begin{equation}
{\cal B}_F(\gamma )=\left(
\begin{array}{cc}
u(\gamma ) &  v(\gamma ) \\
-v(\gamma ) & \;\; u(\gamma )
\end{array}
\right) .  \label{bogfer1}
\end{equation}

The vector $|\gamma \rangle_{ab}$ can be cast in a TFD state. To see
that let us write
\begin{equation}
 |\gamma \rangle_{ab}=[1-\gamma(ba-a^{\dagger}b
^{\dagger})+\frac{\gamma^{2}}{2!}(ba-a^{\dagger}b^{\dagger})^{2}+...]
|0\rangle_{ab} .\label{jot21}
\end{equation}
Using the relations $(ba-a^{\dagger}b^{\dagger})^{2n}|0\rangle_{ab}
=(-1)^{n}|0\rangle_{ab}$ and introducing the reparametrization
$u(\gamma) = \cos\gamma=(1+e^{-\tau})^{-1/2}$, $v(\gamma) =
\sin\gamma=(1+e^{\tau})^{-1/2}$, we obtain
\begin{equation}
|\gamma\rangle_{ab} =
\frac{1}{\sqrt{1+e^{-\tau}}}\,(1+e^{-\tau/2}a^{\dagger
}b^{\dagger})|0\rangle_{ab} .\label{fer21}
\end{equation}
Defining $Z(\tau)=1+e^{-\tau}, $ Eq.~(\ref{fer21}) reads
\begin{equation}
| \gamma \rangle_{ab} =  \frac{1}{\sqrt{Z(\tau)}} \, e^{-\tau N/2}
\left( |0\rangle_a |0 \rangle_b + |1\rangle_a |1 \rangle_b \right),
\end{equation}
where $N=a^{\dagger}a$, the fermion number operator for the mode
$a$, is such that $N|n\rangle_a=n|n\rangle_a $. Therefore, we obtain
\[
|\gamma\rangle=\sqrt{f_{a}(\tau)} \,\sum_{n=0}^{1} |n \rangle_a
|n\rangle_b ,
\]
with $f_{a}(\tau)=Z^{-1}(\tau)\exp(-\tau a^{\dagger}a)$.

With these results, we can prove the following statement. Given the
two fermion displacement operators, $D_a(\alpha)=\exp( a^{\dagger}
\alpha - \alpha^*  a )$ and $D_b(\beta)=\exp( b^{\dagger} \beta -
\beta^* b )$, where $\alpha$ and $\beta$ are Grassmann numbers, then
\begin{equation}\label{squee21}
S_{ab}(\gamma)D_a(\alpha)D_b(\beta) =
D_a(\bar{\alpha})D_b(\bar{\beta})S_{ab}(\gamma)
\end{equation}
where
\begin{equation}
\left(
\begin{array}{c}
\bar{\alpha} \\
\bar{\beta}^{\ast }
\end{array}
\right) = {\cal B}_F(\gamma )\left(
\begin{array}{c}
\alpha \\
\beta^{\ast }
\end{array}
\right) ,  \label{bogfer22}
\end{equation}
Thus, the fermion version of the CS state, defined by
\[
|\alpha,\beta;\gamma\rangle\rangle = S_{ab}(\gamma
)D_a(\alpha)D_b(\beta)|0\rangle_{ab},
\]
is related to the state $|\alpha,\beta;\gamma\rangle  =
D_a(\alpha)D_b(\beta)S_{ab}(\gamma
 )|0\rangle_{ab}$
by the transformation given in Eqs.~(\ref{squee21}) and
(\ref{bogfer22}). As in the bosonic case, when $\alpha=\beta=0$ we
have the two-fermion squeezed vacuum state $|\gamma\rangle_{ab} =
S_{ab}(\gamma) |0\rangle_{ab}$.

Now we turn our attention to the nature of the entanglement in
squeezed fermion states. Considering the states $|\alpha,\beta;
\gamma\rangle$ and inspired by the definition of the density
operator given in Eq.~(\ref{cahill12}), we introduce  the following
density matrix
\[
\rho_{ab}=|-\alpha,-\beta,\gamma\rangle\langle \gamma, \beta, \alpha
|.
\]
Performing the trace in the mode $b$ and using the properties
derived before, we can prove that
$\rho_{a}=D_a(\alpha)f_{a}(\tau)D_a(\alpha)^{\dagger}$, similar to
the boson case. Thus we find that the state $| \alpha, \beta, \gamma
\rangle $ has reduced density operator in the form of a Gibbs-like
density. The reduced entropy is thus maximal. However, in the
fermionic case, the CS states are not in general physical
states~\cite{fer11} since they involve Grassmann variables. The
two-fermion squeezed vacuum state $|\gamma\rangle_{ab}$ is
nevertheless  physical and maximally entangled.

It is worth mention that, in the case of fermions, there is another
class of physical states having maximum entanglement for a given
value of the reduced energy. In fact, one can show that the state
\begin{equation}
|\gamma\rangle^{\prime}_{ab} =  \left( |0\rangle_a |1\rangle_b
+ e^{-\tau/2} |1\rangle_a |0\rangle_b
\right)/{\sqrt{Z(\tau)}}
\end{equation}
has reduced density operator
$\rho^{\prime}_{a}$ identical to the reduced density operator
$\rho_a$ associated with the state $|\gamma\rangle_{ab}$; therefore,
these states have identical reduced energy and entropy.

Summarizing, in this paper we have analyzed a class of two-mode
squeezed boson and fermion states, looking for explicit realization
of maximum entangled states with fixed energy.  We investigate the
case of bosons, and then, construct the fermion version, to show
that such states, in both cases, are  maximum entangled. For
achieving these results we have demonstrated some relations
involving the squeezed boson states, which are  then extended to the
case of fermions. The calculations for fermions are performed  with
a generalization of the  density fermion operator introduced by
Cahill and Glauber~\cite{fer11}.

This work was partially supported by CNPq (Brazil) and NSERC
(Canada).

\end{document}